\documentclass[aps,pra,twocolumn,showpacs]{revtex4}
\usepackage[dvips]{graphics,graphicx}
\usepackage{graphicx,epsfig}
\usepackage[usenames,dvipsnames]{color}

\begin{document}

\title{Quantum particle in a parabolic lattice in the presence of a gauge field} 
\author{Andrey R. Kolovsky}
\affiliation{Kirensky Institute of Physics, 660036 Krasnoyarsk, Russia}
\affiliation{Siberian Federal University, 660041 Krasnoyarsk, Russia}
\author{Fabian Grusdt$^1$ and Michael Fleischhauer}
\affiliation{Dept. of Physics and research center OPTIMAS, University of Kaiserslautern, 67663 Kaiserslautern, Germany} 
\affiliation{$^1$ Graduate School Materials Science in Mainz, 67663 Kaiserslautern, Germany} 

\begin{abstract}
We analyze the eigenstates of a two-dimensional lattice with additional harmonic confinement in the presence of an artificial magnetic field. While the softness of the confinement makes a distinction between bulk and edge states difficult, the interplay of harmonic potential and lattice leads to a different classification of states in three energy regions: In the low-energy regime, where lattice effects are small, all states are transporting topologically non-trivial states. For large energies above a certain critical value, the periodic lattice causes localization of all states through a mechanism similar to Wannier-Stark localization. In the intermediate energy regime transporting, topologically non-trivial states coexist with topologically trivial counter-transporting chaotic states. The character of the eigenstates, in particular their transport properties are studied numerically and are explained using a semiclassical analysis.
\end{abstract}

\maketitle

\section{Introduction}
\label{sec0}

Ultra-cold atoms in optical lattices have established themselves as
powerful model system offering unique experimental facilities 
for studying many fundamental phenomena of solid-state and many-body physics.
E.g. the near absence of dissipation in optical lattices has lead to the 
observation of single-particle quantum interference effects, such as Bloch oscillations or 
Landau-Zener tunneling \cite{Daha96,Hall10,Zene09}, as well as
Anderson localisation in a disorder potential \cite{Roat08}. The ability to tune 
interactions e.g. by spatial confinement made it possible to drive interaction 
induced quantum phase transitions in cold atom experiments \cite{Grei02} and the
variety of lattice geometries possible allows to observe e.g. magnetic frustration
in triangular lattices \cite{Stru11}.

Particularly interesting in this context is the recent experimental realization of artificial magnetic 
fields in lattices \cite{Aide11,Stru12,Aide13,Miy13} that opens prospects for studying quantum Hall effects and Chern insulators with neutral atoms. To understand how well this system can reproduce the solid-state Hall physics and what novel effects may arise in the cold-atom setting, 
it is important to understand the role of boundary conditions that principally differ from the Dirichlet boundary conditions in solid crystals. This problem was addressed recently in \cite{Buch12,Gold12,Gold13b,Gold13}, where a quantum particle in a 2D square lattice subject to an Abelian gauge field was considered and the effects of a confinement potential $V({\bf r}) \sim (x^\delta+y^\delta)$ \cite{Buch12} and  $V({\bf r}) \sim r^\delta$  \cite{Gold12,Gold13b,Gold13} 
was studied. These potentials impose smooth boundaries with a variable steepness characterized by the parameter $\delta$. It was shown that there is no principle difference between Dirichlet ($\delta=\infty$) and smooth boundaries if  $\delta\ge4$. For $\delta\ge 4$ one can clearly distinguish edge states from bulk Landau states which is believed to be a precondition to mimic solid-state Hall physics with cold atoms. The case $\delta=2$, which is typically realized in laboratory experiments, appeared to be more subtle, with no clear conclusions and contradictory statements that separation between the edge and bulk states is possible \cite{Gold13b} or not \cite{Buch12}.

\begin{figure}[b]
\center
\includegraphics[width=0.99 \columnwidth]{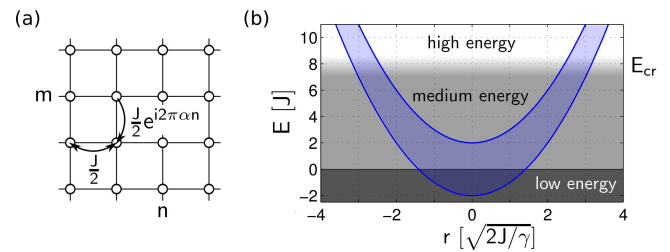}
\caption{(Color online) {\it left:} Two-dimensional square lattice model with complex hopping amplitudes
equivalent to a magnetic field perpendicular to the lattice. There is an additional harmonic confinement potential as is the case for most cold-atom experiments. The Peierls phase $\alpha$ quantifies the
flux per plaquette in units of the flux quantum. {\it right:} 
Shaded region between parabolas (blue): sketch of allowed spatial and energy regions for particle trajectories in local density approximation and for $\alpha=0$. Characteristic energy scales that will be introduced in the paper are the minimum energy $E\approx-2J$ (lower solid parabola), the energy $E=0$ where chaotic trajectories appear (central dashed parabola) and $E_{\text{cr}}$ where a delocalization-localization crossover takes place.}
\label{fig1}
\end{figure}

The aim of the present work is a detailed analysis of the spectrum and eigenstates of a quantum particle (an atom) in a 2D lattice with harmonic confinement (so-called parabolic lattice \cite{remark1}) in the presence of an artificial magnetic field (see Fig.~\ref{fig1}). In other words, we focus on the above mentioned case of $\delta=2$. Without lattice potential, the eigenstates are identical to Landau levels (LL) in symmetric gauge. 
If the harmonic confinement is weak, such that the oscillator frequency is small compared to the cyclotron frequency,  every eigenstate with energy $E$ 
corresponding to lowest LL states
is localized at the boundary of the classically allowed region fixed by $E$, and it makes no sense to distinguish
edge and bulk states. Numerical simulations of eigenstates and spectra, of wavepacket dynamics and the effect of a local flux insertion show, however, 
that the presence of a lattice potential gives rise to a different classification of eigenstates
in three regimes: a low-, medium- and high-energy regime. 
The structure of the corresponding eigenstates will be explained making use of a semiclassical analysis as well as recent analytical results about Landau-Stark states  \cite{85,90}, which are 
eigenstates of a quantum particle in the presence of a `magnetic' field normal to the lattice plane  and an in-plane `electric' field. We will introduce a classification of quantum states into topologically non-trivial transporting states, chaotic counter-transporting states and localized states. We also identify a new fundamental frequency [the encircling frequency, see Eq.~(\ref{b3}) in Sec.~\ref{sec2}], which takes over the role of cyclotron frequency for the quantum particle in a plane lattice.

\section{model}
\label{sec1}

We consider neutral atoms in a a two-dimensional square lattice with period $a=1$ in the tight-binding limit, as indicated in Fig.\ref{fig1}.
The atoms are subject to an artificial magnetic field and there is an additional harmonic confinement. Using the Landau gauge the corresponding Hamiltonian reads 
\begin{eqnarray}
&&(\widehat{H}\psi)_{n,m} = -\frac{J}{2}\left(e^{i2\pi\alpha n}\psi_{n,m+1} + e^{-i2\pi\alpha n}\psi_{n,m-1}\right)\nonumber
\\
&& -\frac{J}{2}\left(\psi_{n+1,m}  +  \psi_{n-1,m}\right)
+\frac{\gamma}{2}(n^2+m^2)\psi_{n,m} \;, \label{a0}
\end{eqnarray}
where $n$ and $m$ label the sites of the square lattice, $J$ is the hopping matrix element, and $\gamma$ the strength of harmonic confinement.
(The latter parameter can be expressed through the trap frequency $\omega_{\text{hc}}$ and the atom mass $M$ as $\gamma=Ma^2\omega_{\text{hc}}^2$.) The presence of the magnetic field is encoded in the 
Peierls phase $\alpha$ which is equal to the magnetic flux per plaquette in units of the flux quantum. 
In order to be close to Landau-level physics in a homogeneous system, we will assume throughout this paper that the magnetic unit cell is much larger than the lattice unit cell.   
In most cases  we use $\alpha=1/6$ or $1/10$.

\section{numerical simulation of the quantum problem}
\label{sec2}

\subsection{Eigenstates and spectrum}
\label{subsec2-1}

If $\gamma=0$ the spectrum of (\ref{a0}) consists of a finite (rational $\alpha$) or infinite (irrational $\alpha$) number of magnetic sub-bands in the energy interval $-2J<E<2J$. The presence of the harmonic confinement changes this spectrum. It is now unbounded from above, $-2J<E<\infty$, and one finds only remnants of the magnetic sub-bands in the form of steps in the mean density of states (DOS) for $E<2J$ \cite{Gerb10}.  With increase of the energy above $2J$ the mean density of states approaches the value
\begin{equation}
\label{density}
\rho(E)=2\pi/\gamma \;,
\end{equation}
which coincides with the density of states of (\ref{a0}) for $\alpha=0$. 
Although $\rho(E)$ for $\alpha\ne0$ looks similar to that for $\alpha=0$, the details of the spectrum and the eigenstates are completely different. 

If $\gamma$ is small and  $\alpha=0$, a qualitative picture of the effect of the harmonic confining potential can be obtained by treating it as a space-dependent
chemical potential, which leads to the overall band structure depicted in Fig.~\ref{fig1}. 
As indicated in
the figure we will show in this paper that one has to distinguish three qualitatively different energy regions:
a low-energy regime $E<0$, a high-energy regime $E>E_{\rm cr} > 2J$  and a medium energy regime in between.
The critical energy will be shown to be 
\begin{equation}
E_{\rm cr} = \frac{(2\pi \alpha J)^2}{2\gamma}=E_R\left(\frac{\omega_\alpha}{\omega_{\text{hc}}}\right)^2 \;,
\label{eq:EcrIntro}
\end{equation}
where $E_{R}=\hbar^2/Ma^2$ is proportional to the recoil energy and $\omega_\alpha=2\pi\alpha J/\hbar$ has the meaning of the cyclotron frequency. As will be seen in the following,  the nature of the eigenstates is very different in the different parts of the spectrum. 

\begin{figure}[htb]
\center
\includegraphics[width=0.9 \columnwidth]{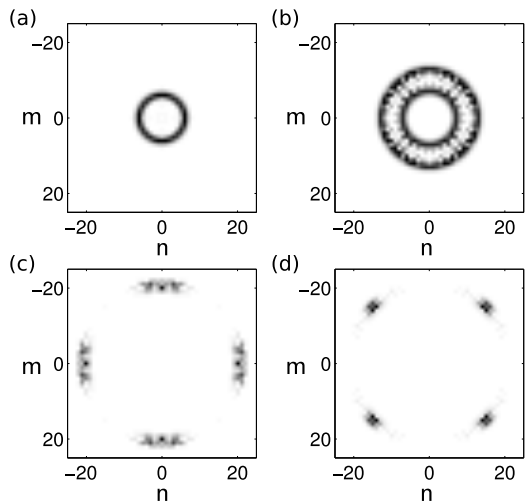}
\caption{The 25th (a),  349th (b), 1365nd (c), and 1370th (d) eigenstates of the Hamiltonian (\ref{a0}) with energies 
$E/J=-0.5511, 2.7671, 10.8544, 10.9069$, respectively. The squared absolute value of the wave-function 
$\Psi_{n,m}$ is shown as a gray-scale map. We used $\alpha=1/6$ and $\gamma/J=0.05$.}
\label{fig-7}
\end{figure}

Numerical diagonalization of the Hamiltonian eq.(\ref{a0}) allows us to easily calculate the single-particle spectrum and the corresponding 
eigenstates of the problem.
In Fig.~\ref{fig-7}  we have plotted the absolute square of the eigenstates for energy values corresponding to the different regions. One notices a qualitative change in the character of the states when increasing the energy. For low energies, as shown in Fig.~\ref{fig-7}a ($E/J=-0.6530$),
the eigenstates are circular, resembling Landau states with fixed angular momentum. For intermediate energies, as shown in Fig.~\ref{fig-7}b ($E/J=2.4725$)  
there is a more complex pattern. There are two circular structures, an inner and an outer one, corresponding
to the boundaries of the classically allowed regions for the given energy (see Fig.~\ref{fig1}b). In addition there is however also a finite probability 
amplitude in the spatial region between the two rings. Finally when going to high energies, as shown in Figs.~\ref{fig-7}c and \ref{fig-7}d
localized structures emerge with a four-fold symmetry corresponding to the underlying square lattice.

\subsubsection{Low-energy region}
\label{subsubsec:lowEnergyStates}

In the low-energy part of the spectrum we can use the long-wavelength- or effective mass approximation to write the
Hamiltonian, eq.(\ref{a0}), in the form (here and in the following we set $\hbar =1$)
\begin{equation}
\label{c5}
\widehat{H}=\frac{1}{2M^*}\left(\hat{\bf{p}}-{\bf A}\right)^2
+\frac{\gamma}{2}(x^2+y^2) .
\end{equation}
Here $M^*= J^{-1}$ is the effective mass, and ${\bf A}= 2 \pi \alpha (-y/2,x/2)$ the vector potential (where we have changed to the symmetric gauge).  It is convenient  to express (\ref{c5}) in the  form
\begin{equation}
\label{c6}
\widehat{H}=\widehat{H}_0 - \omega\widehat{L}_z  \;,\quad
\widehat{H}_0=\frac{\hat{\bf{p}}^2}{2M^*} +\frac{\gamma+M^*\omega^2}{2}(x^2+y^2) \;,
\end{equation}
where $\widehat{L}_z$  is the angular-momentum operator, and $\omega=J\pi\alpha$. The eigenstates of (\ref{c6}) are labeled by the angular-momentum, $n_L$, and the radial quantum number, $n_r$, defining the Landau level \cite{Darwin1931}. 
These quantum numbers can be used to characterize the low-energy eigenstates of the system (\ref{a0}) in spite of the fact that, strictly speaking, the eigenstates do not possess rotational symmetry due to the lattice potential.

\begin{figure}
\center
\includegraphics[width=0.9 \columnwidth]{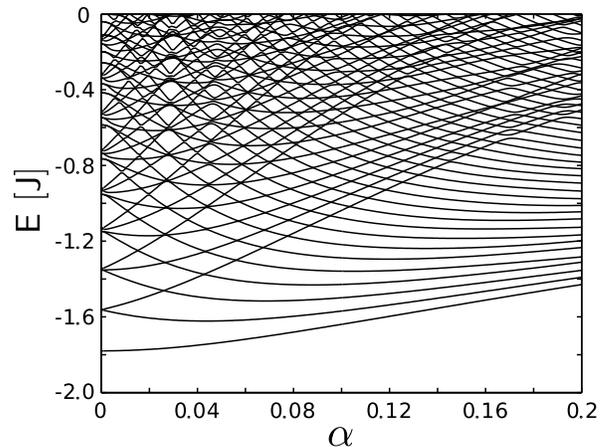}
\caption{The low-energy part of the spectrum for $J=1$ and $\gamma =0.05$ as a function of the Peierls phase $\alpha$.}
\label{fig11}
\end{figure}

Fig. \ref{fig11} shows the lower part of the energy spectrum of (\ref{a0}) as the function of the Peierls phase $\alpha$, i.e. the magnetic flux per plaquette. One clearly sees the Zeemann splitting of degenerate levels of the 2D harmonic oscillator,  where the first level has quantum numbers $(n_r,n_L)=(0,0)$,  the second level $(0,1)$ and $(1,-1)$, the third $(0,2)$, $(1,0)$, $(2,-2)$, the 4th $(0,3)$, $(1,2)$, $(2,-1)$, $(3,-3)$, the 5th $(0,4)$, $(1,2)$, $(2,0)$, $(3,-2)$, $(4,-4)$, and so on. With increase of $\alpha$ these levels rearrange in a pattern which consists of series of levels with fixed $n_r$ and monotonically increasing $n_L$. 

A different interpretation of the low-energy spectrum is based on the observation that  for $\gamma=0$  the Hamiltonian (\ref{c6}) defines the degenerate Landau levels with level spacing given by the cyclotron frequency 
\begin{equation}
\label{c7}
\omega_\alpha=2\pi\alpha J=2\omega \;, \quad |\alpha|\ll 1/2 \;.
\end{equation}
The harmonic confinement splits these degenerate levels into series of levels with equidistant spacing $\Omega$. For small $\gamma$ one finds 
\begin{equation}
\label{b3} 
\Omega=\frac{\gamma}{2\pi\alpha} \;.
\end{equation}

\subsubsection{medium energy region}

\begin{figure}
\center
\includegraphics[width=0.9 \columnwidth, clip]{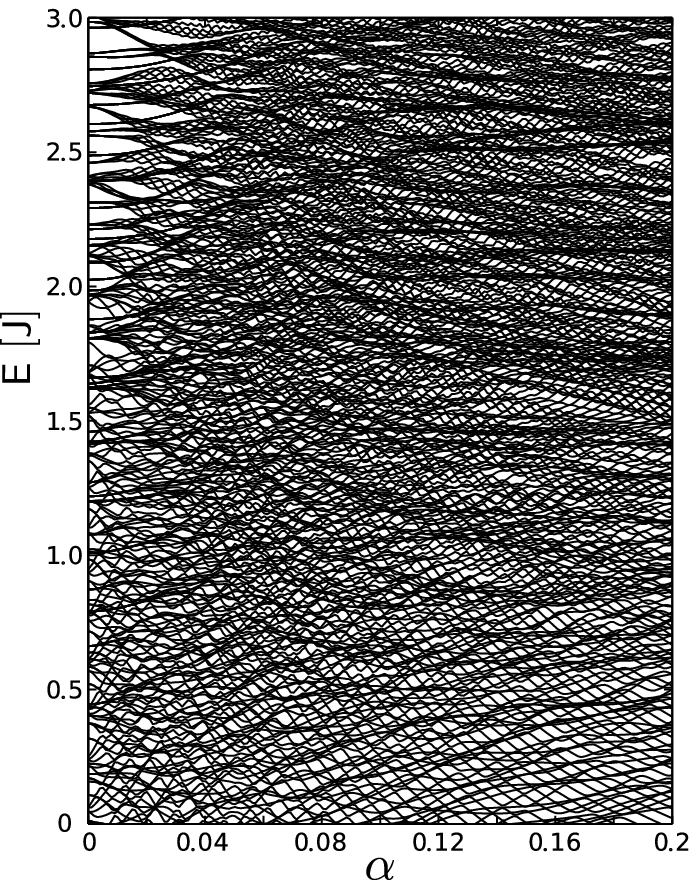}
\caption{A fragment of the medium energy spectrum of the Hamiltonian (\ref{a0}) for $J=1$, $\gamma=0.05$, and $0\le\alpha\le0.2$.}
\label{fig6}
\end{figure} 

Figure \ref{fig6} shows a fragment of the medium energy spectrum of Hamiltonian (\ref{a0}) as function of the Peierls phase $\alpha$ for $J=1$ 
and $\gamma=0.05$. Here a rather complicated level pattern is noticed. There is a large number of levels with true or  nearly avoided crossings.
One notices, however, that some regular structures of the low-energy spectrum survive.

\subsubsection{High-energy region}

\begin{figure}[htb]
\center
\includegraphics[width=0.9 \columnwidth, clip]{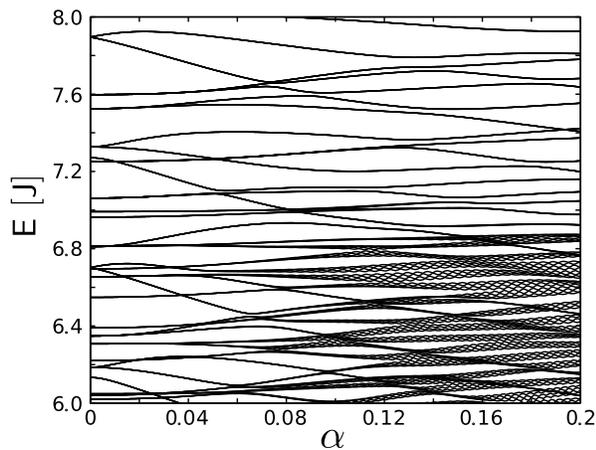}
\caption{A fragment of the high-energy spectrum of the Hamiltonian (\ref{a0}), for $J=1$, $\gamma=0.05$, and $0\le\alpha\le 0.2$}
\label{fig8}
\end{figure}

The characteristic feature of the high-energy spectrum above a certain critical value $E_{\rm cr}$, a fragment of which is shown in Fig.~\ref{fig8}, is an approximate four-fold degeneracy of the energy levels.  This degeneracy reflects localization of the eigenstates in segments of the circles, see Fig.~\ref{fig-7}(d). Due to the 4-fold lattice symmetry there are three other eigenstates with almost the same energy which look similar to the depicted state. From this set of four exact states one can construct a new set of four approximate eigenstates, where every state is localized only in one segment. Thus a particle  with the mean energy $E>E_{\rm cr}$, which is initially localized within one
of the segments, remains localized in this segment for exponentially large times. 

\subsection{Dynamical signatures of localization}

\begin{figure}[htb]
\center
\includegraphics[width=0.7 \columnwidth]{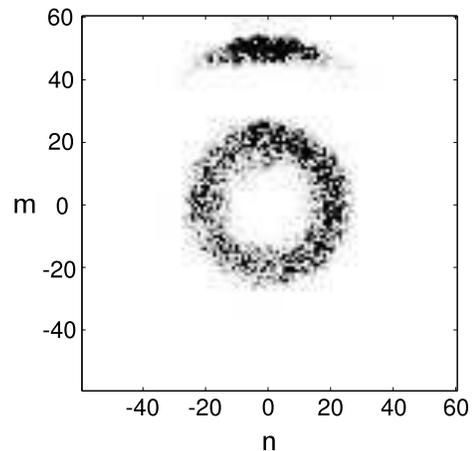}
\caption{Populations of the lattice sites at the end of numerical simulation. Initial conditions corresponds to a narrow scrambled Gaussian centered at $(n_0,m_0)=(0,20)$, case (a), and $(n_0,m_0)=(0,50)$, case (b). Parameters are $J=1$, $\alpha=0.1$, and $\gamma=0.01$. Evolution time is $t=30T_J$
, ($T_J =2\pi /J$).}
\label{fig9}
\end{figure}

To experimentally observe the characteristics of eigenstates at different energies, we suggest to
measure dynamics of an initially localized wave-packet.
Our numerical experiment follows the scheme of laboratory experiments on dipole oscillations of cold atoms in parabolic lattices. 
The protocol involves  a sudden shift of the origin of the harmonic potential by distance $r_0$, so that the atomic cloud appears on the slope of the parabolic lattice where it has the energy $E\approx\gamma r_0^2/2$. Then the system  evolves freely for a certain time $t$, which is followed by (destructive) measurement of the atomic density. The result of this numerical experiment is shown in Fig.~\ref{fig9}, where we have chosen a narrow Gaussian distribution with random phases
as initial condition. The packet is shifted from the lattice origin by $m_0=20$ sites, case (a), and $m_0=50$ sites, case (b). In the former case the packet energy is smaller than the critical energy $E_{\rm cr}$ and it encircles the lattice origin. In the latter case, where $E>E_{\rm cr}$, the wave packet remains localized. Since the wave-packet dynamics can be expressed in terms of eigenstates, this result undoubtedly indicates the qualitative difference between 
states below and above $E_{\rm cr}$.

\subsection{Effects of disorder}
An important characteristics of topological systems is their robustness against disorder. Because of 
their effective description in terms of the topologically non-trivial lowest Landau level states (c.f. sec.\ref{subsubsec:lowEnergyStates}), 
we expect low-energy states to be robust to such disorder. In the following we will show that this is indeed the case,
while high-energy states are very sensitive to disorder.

\begin{figure} [htb]
\center
\includegraphics[width=0.8 \columnwidth]{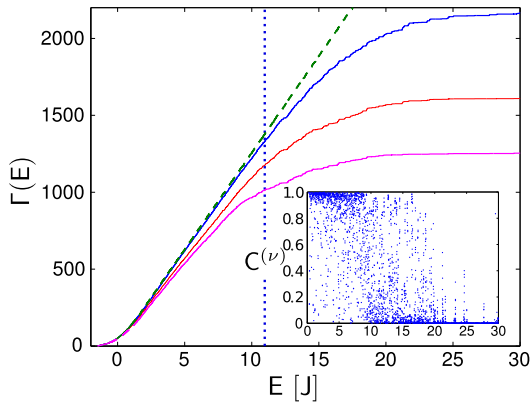}
\caption{(Color online) {\it main panel:} Effect of disorder on the eigenstates. Shown is the cumulative sum $\Gamma(E) = \sum_\nu C^{(\nu)}$ for different disorder strength $\epsilon/J=0.01$, $0.05$, $0.1$, solid lines from top to bottom. The vertical (blue) dotted line indicates the critical energy and the (green) dashed line shows the
integrated DOS in the absence of disorder. {\it Inset:} The quantity (\ref{c4}) for a specific realization of a weak on-site random potential $|V_{n,m}|\le \epsilon/2$ with $\epsilon=0.1 J$. The eigenstates are ordered by their energies $E$. We used $\alpha=1/6$ and $\gamma/J=0.05$.}
\label{fig10}
\end{figure}

To this end we analyze the robustness of these states by adding a weak random on-site potential $V_{nm}$ to the Hamiltonian (\ref{a0}) in our numerical simulations. $V_{nm}$ is a Gaussian distribution with vanishing mean value and is assumed to be spatially uncorrelated
\begin{equation}
 \left[\overline{\bigl. V_{nm} V_{n^\prime m^\prime}\bigr.}\right]^{1/2} = \epsilon\, \delta_{n,n^\prime} 
\delta_{m, m^\prime}.
\end{equation}
Here $\epsilon$ describes the strength of the disorder potential. A convenient characteristics of the robustness of an eigenstate $\Psi^{(\nu)}$ to disorder is the quantity
\begin{equation}
\label{c4}
C^{(\nu)}=\biggl.\Bigl|\sum_{n,m} \Psi_{-n,-m}^{(\nu)}(\Psi_{n,m}^{(\nu)})^*\Bigr|\biggr. \le 1 \;. 
\end{equation}
where the sum runs over all lattice sites. 
For a vanishing random potential one can prove using eq. (\ref{a0}) that all eigenstates are symmetric or antisymmetric functions with respect to reflection $n\rightarrow -n$ and $m\rightarrow -m$, i.e.,
\begin{equation}
\label{c3}
\Psi_{-n,-m}=\pm \Psi_{n,m} \;.
\end{equation}
In this case $C^{(\nu)}$ equals to unity and 
\begin{equation}
\Gamma(E)=\sum_{\nu, E_\nu\le E} C^{(\nu)}
\end{equation}
is equal to the integrated density of states up to energy $E$.

Fig.~\ref{fig10} shows $\Gamma(E)$ for increasing disorder strength as well as the corresponding
function in the absence of disorder (dashed  curve). 
The random potential breaks the symmetry (\ref{c3})
and $C^{(\nu)}$ quickly drops below unity, see inset of Fig.~\ref{fig10},  where the 
eigenstates are sorted according to their energies.
One notices a remarkable agreement of $\Gamma(E)$ with its value in the absence of disorder in the low-energy region. 
The deviations from the disorder-free curve increase with increasing $\epsilon$ but stay small until the critical energy $E_{\rm cr}$.
Thus the random potential affects only some of the eigenstates for $E<E_{\rm cr}$.
Above $E_{\rm cr}$, however, $\Gamma(E)$ saturates which indicates that the disorder completely randomizes the eigenstates.
We conclude that most of the states with $E_\nu<E_{\rm cr}$ are robust against impurities ($C^{(\nu)}\approx 1$) while most of the states with 
$E_\nu>E_{\rm cr}$ are not ($C^{(\nu)}\approx 0$). Let us emphasize that at $E_{\rm cr}$ a crossover takes place, rather than a sharp transition.
 
\

In the following sections we want to provide some understanding of the spectrum, the eigenstates and their transport properties for $\alpha\ne0$.

\section{Classical approach}
\label{sec3}

Following the line of Ref.~\cite{85,90} we provide here a classical analysis of the system (\ref{a0}), which allows insight into the
transport properties of the states. The classical counterpart of the quantum Hamiltonian (\ref{a0}) reads 
\begin{equation} 
\label{b0}
H_{cl}=-J\cos(p_x) - J\cos(p_y - \tilde{x}) +\frac{\tilde{\gamma}}{2}(\tilde{x}^2+\tilde{y}^2)  \;,
\end{equation}
where $\tilde{\gamma}=\gamma/(2\pi\alpha)^2$.  As shown in \cite{90} the classical limit corresponds to $\alpha\rightarrow0$ while keeping $\tilde{\gamma}$ constant,  and in this case $2 \pi \alpha n \rightarrow \tilde{x}$ and $2 \pi \alpha m \rightarrow \tilde{y}$ become continuous variables. 
Thus the only relevant parameter of the classical dynamics is $\tilde{\gamma}/J$ \footnote{Note that, like the critical energy $E_{\text{cr}}$ from eq. (\ref{eq:EcrIntro}) 
in the quantum case, the classical parameter $\tilde{\gamma}/J = \frac{J}{2 E_R} \left( \frac{\omega_{\text{hc}}}{\omega_{\alpha}} \right)^2$ is determined by the ratio between cyclotron- and trap frequency.}. In what follows, however, we shall use a different form of the classical Hamiltonian,
\begin{equation}
\label{b1} 
H_{cl}=-J\cos(p_x) - J\cos(p_y-2\pi\alpha x) +\frac{\gamma}{2}(x^2+y^2)  \;,
\end{equation}
which is obtained from the previous Hamiltonian by obvious scaling of the coordinates $x$ and $y$. This form allows a direct comparison of the classical trajectories with the quantum wave functions for a finite $\alpha$.

\begin{figure}
\includegraphics[width=0.9 \columnwidth,clip]{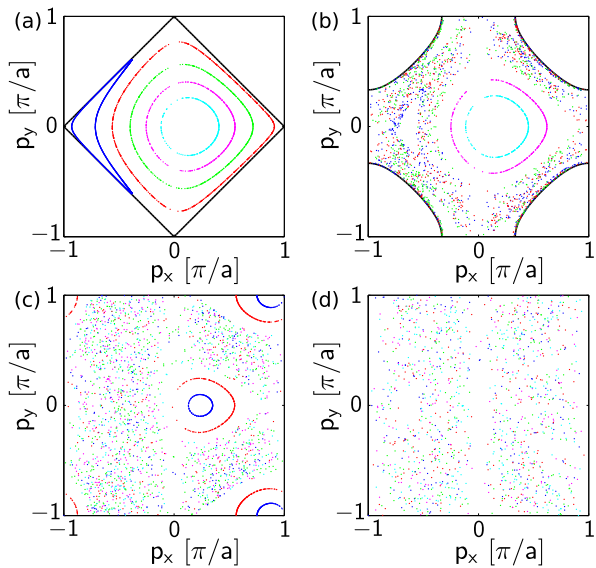}
\caption{(Color online) Poincare cross-sections by the plane $\vartheta=0$ for energy $E/J = 0, 0.5, 3, 11$ (a-d). The system parameters are $\alpha=1/6$ and $\gamma/J=0.05$. The solid lines in the panels (a) and (b) restrict  the available phase space.}
\label{fig3}
\end{figure}

For a given energy $E$ any phase trajectory of (\ref{b1}) is uniquely described by momenta $p_x$ and $p_y$ and the angle $\vartheta=\arctan(x/y)$, which are cyclic variables. Thus the energy shell of (\ref{b1}) lies inside the three-dimensional torus or coincides with this torus if $E\ge 2J$. Fig. \ref{fig3} shows the Poincare cross-sections 
of the energy shell by the plane $\vartheta=0$ for few values of $E$. 
While for  $E<  0$ we only found regular  trajectories, for $E\ge 0$, where the effective mass approximation
(\ref{c5}) fails completely due to the lattice,
the typical structure of a non-integrable system with mixed phase space is noticed. 

\begin{figure}
\includegraphics[width=0.7 \columnwidth, clip]{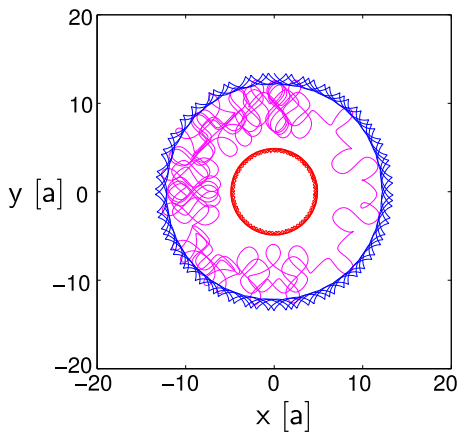}
\caption{(Color online) Examples of classical trajectories with $E/J=2.5$ for three different initial conditions: $(p_x,p_y)\approx(0,0)$ (outer, blue), $(\pi,\pi)$ (inner, red), and $(\pi/2, \pi/2)$ (middle, magenta). The initial value of $x$ is zero for all three trajectories and the initial value of $y$ is adjusted to ensure equal energies. The evolution  time corresponds to 5 periods of frequency (\ref{b3}). We used parameters $\alpha=1/6$ and $\gamma/J=0.05$.}
\label{fig4}
\end{figure}

\begin{figure}
\center
\includegraphics[width=0.8\columnwidth, clip]{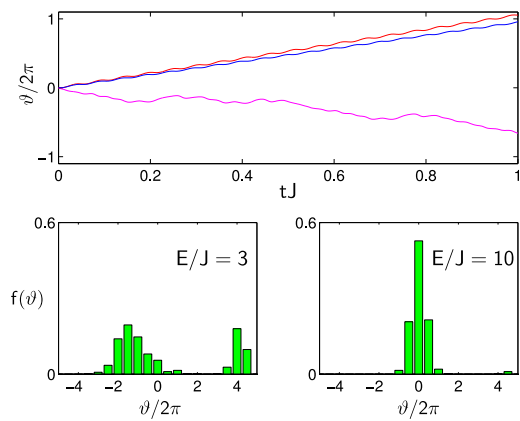}
\caption{(Color online) {\it Upper panel:} Winding angle $\vartheta=\arctan(x/y)$ as the function of time for the 3 trajectories shown in Fig.~  \ref{fig4}. Upper curve is the inner regular trajectory, middle curve the outer regular trajectory, and the lower curve the chaotic trajectory.
{\it Lower panels:} Distribution function $f(\vartheta,t)$ for the winding angles $\vartheta$  at $t=4T_\Omega = 8 \pi / \Omega$ for 400 different trajectories on the energy shell $E/J=3$ (left) and $E/J=10$ (right).
}
\label{fig5}
\end{figure}

Let us first discuss the case of moderate energy, where the phase space consists of two big stability islands surrounded by the chaotic sea
(see Fig.~\ref{fig3}c). 
For reasons that become clear below, we will refer to them as inner and outer transporting islands.
The blue and red lines in Fig.~\ref{fig4} show the particle trajectories for initial conditions inside the central and lower/upper stability islands, respectively. Additionally, the winding angle  $\vartheta=\arctan(x/y)$ is depicted in Fig.~\ref{fig5}(a) as function of time. It is seen from Fig.~\ref{fig5}
that both of these trajectories encircle the coordinate origin clockwise with frequency $\Omega$ given in Eq.~(\ref{b3}). 
We will refer to them as \emph{transporting} trajectories.

In the classical approach the encircling frequency is determined by the drift velocity of a charged particle in the crossing electric and magnetic fields. In fact, locally we can approximate  the parabolic potential by a gradient force pointing to the coordinate origin,
\begin{equation}
\label{b2}
{\bf F}=-\gamma {\bf r} \;,\quad {\bf r}=(x,y) \;.
\end{equation}
In the notation used throughout the paper, the drift velocity in the direction perpendicular to ${\bf F}$ is given by \cite{85}
\begin{equation}
\label{a4} 
v^*=F/2\pi\alpha \;.
\end{equation}
Thus the encircling period is $T_\Omega\equiv 2\pi/\Omega=2\pi r/v^*= (2\pi)^2 \alpha/\gamma$.  

Closer inspection of the numerical data in  Fig.~\ref{fig5}(a) shows that the encircling frequency for the outer trajectory in Fig.~\ref{fig4} is slightly smaller than $\Omega$, while for the inner trajectory it is slightly larger. 
More importantly the inner trajectories appear only when $E$ approaches $2J$.
Let us mention that angular momentum is an approximate integral of motion for the transporting trajectories as 
$\dot\vartheta \approx$ const. and $r\approx$ const. (see Figs. ~\ref{fig4} and \ref{fig5}(a)), and as 
\begin{equation}
\dot{\vartheta}=\frac{L}{r^2} \;,\quad L=\dot{{\bf r}}\times{\bf r} \;.
\end{equation}
%

If the initial condition is chosen outside the transporting islands (magenta lines in Figs.~\ref{fig4}, \ref{fig5}(a)), yet another type of trajectory is found. These trajectories appear to be chaotic, and in average rotate counter-clockwise. We will refer to them as \emph{counter-transporting} trajectories.
Because of their chaotic nature, they do not have a well defined encircling frequency. 
Instead we find numerically for sufficiently large energies that
they obey the so-called summation rule. Namely, if we chose an ensemble of classical particles uniformly distributed over the energy shell, the clockwise current due to (regular) transporting trajectories and the counter-clockwise current due to (chaotic) counter-transporting trajectories compensate each other. We conjecture that this sum rule is valid as soon as the entire phase space is classically allowed, although we are not aware of an exact proof. 

The sum rule is also illustrated in the lower panels of Fig.~\ref{fig5} which show the distribution of the winding angles $\vartheta$ at $t=4T_\Omega$ for 400 different trajectories with random initial conditions, yet $x(t=0)=0$, and fixed energy $E/J=3$ (left panel) and $E/J=10$ (right panel). The double-peak structure of the distribution function $f=f(\vartheta,t)$ reflects the presence of both transporting and counter-transporting trajectories. It is also seen in the figure that the right peak of $f(\vartheta,t)$, which is associated with transporting trajectories, decreases if the energy is increased. This is due to the decrease of the size of transporting islands in Fig.~\ref{fig3}, which are smaller for larger energy  (i.e., larger gradient force) and completely disappear for energies above the critical energy $E_{\rm cr}$ (see Fig.~\ref{fig3}d). In this case the distribution function $f(\vartheta,t)$ becomes localized within the interval $|\vartheta|<2\pi$, indicating the absence of transport in the system. 

We can obtain an estimate for the critical energy $E_{\rm cr}$ by drawing an analogy with the related problem of Landau-Stark states which, by definition, are eigenstates  of a charged particle in a two-dimensional plane lattice in the presence of in-plane electric field and normal-to-the-lattice  magnetic field. It was shown in  \cite{85,90} that these states show a delocalization-localization crossover at  
\begin{equation}
\label{a3}
F_{\rm cr}=2\pi\alpha J \;.
\end{equation}
Namely, the localization length $\xi_\perp$ of the Landau-Stark states in the direction orthogonal to ${\bf F}$ blows up exponentially when the electric field decreases below the critical value (\ref{a3}) \cite{remark2}. Associating $F$ in (\ref{a3}) with the gradient force (\ref{b2}) and neglecting the kinetic energy in Eq.~(\ref{b1}) we find
\begin{equation}
\label{b4}
E_{\rm cr}\approx(2\pi\alpha J)^2/2\gamma \;. 
\end{equation}
We also mention that the classical counterpart of the quantum Hamiltonian of Landau-Stark states has quite a similar structure of phase space, containing two chains of transporting islands. In this sense, the physics behind localization of the Landau-Stark states and wave-function localization discussed in Sec.~III.A is the same and is related to disappearance of the transporting islands at $F=F_{\rm cr}$.

\section{Topological classification of the quantum states}
\label{sec5}

In the previous section we introduced the notions of transporting and counter-transporting 
trajectories in the classical version of our model and identified them with the various
quantum states obtained in section \ref{subsec2-1}. Now we will investigate the topological properties 
of these states. To this end we reverse Laughlin's argument \cite{Laughlin1981} for the quantization of 
the Hall conductance: we adiabatically introduce local magnetic flux and investigate the system's
response, i.e. the Hall current. This allows us to distinguish between topologically trivial and non-trivial
eigenstates.

\subsection{Local flux}

A local flux insertion at the origin leads to an additional contribution to the vector potential 
\begin{equation}
\label{d1}
{\bf A}(x,y) \sim \left(-\frac{y}{x^2+y^2},\frac{x}{x^2+y^2}\right). 
\end{equation}
It corresponds to a magnetic field which is zero everywhere except at the coordinate origin. 
It is easy to show that the tight-binding counterpart of  (\ref{d1}) results in an additional
contribution to the Peierls phase by 
\begin{eqnarray}
\label{d2}
&&\phi_x(n,m)\equiv \phi(n,m) = \\
&&\quad=\theta \left[ \arctan\left(\frac{2n-1}{2m-1}\right) - \arctan\left(\frac{2n+1}{2m-1}\right) \right] \nonumber
\end{eqnarray}
in the $x$-direction and  $\phi_y(n,m)=\phi(m,n)$ in the $y$-direction. 
Here $\theta$ quantifies the amount of the flux inserted in units of the magnetic flux quantum. 
Thus we have
\begin{eqnarray}
(\widehat{H}\psi)_{n,m}&=& 
-\frac{J}{2}\left(e^{-i\pi\alpha m} e^{-i\phi(n,m)} \psi_{n+1,m}  +  h.c.\right)\nonumber\\
&&-\frac{J}{2}\left(e^{i\pi\alpha n}e^{i \phi(m,n)} \psi_{n,m+1}+ h.c.\right) \label{d3}\\
&&+\frac{\gamma}{2}\left[(n-1/2)^2+(m-1/2)^2\right]\psi_{n,m} \nonumber
\end{eqnarray}
Notice that in comparison with (\ref{a0}) we here used the symmetric gauge for the uniform magnetic field and shift the coordinate origin from the site 
$(n,m)=(0,0)$ to the center of the plaquette.

\begin{figure}
\includegraphics[width=0.95 \columnwidth]{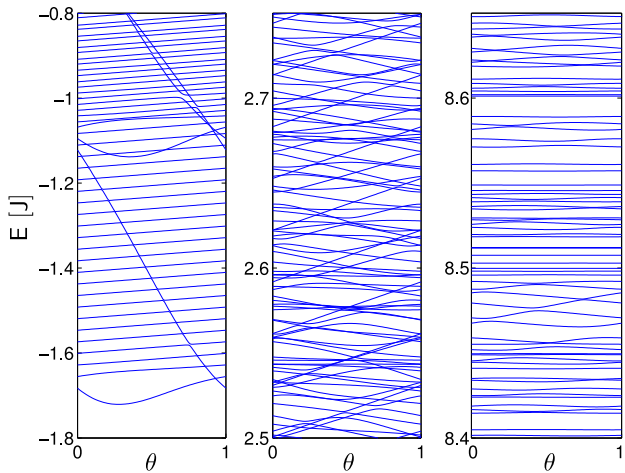}
\caption{(Color online) Different regions of the energy spectrum of Hamiltonian (\ref{d3}) as function of the flux $\theta$ inserted at the origin. Parameters are $J=1$, $\alpha=0.1$ and $\gamma=0.018$. Shown are the low- (right), medium- (middle), and high-energy regions (right). Notice the different scales of the energy axis in the panels.}
\label{fig12}
\end{figure}

Figure \ref{fig12} shows the low-, medium-, and high-energy parts of the spectrum of (\ref{d3}) as function of the inserted flux $\theta$. 
The system parameters  are $J=1$, $\alpha=0.1$ and $\gamma=0.018$, which correspond to the same value of the parameter $\tilde{\gamma}$ that was used in Sec.~\ref{sec2} and Sec.~\ref{sec3} (thus also $E_{\text{cr}}/J$ is the same as in the previous sections). It is seen from  Fig.~\ref{fig12} that the positions of the energy levels for $\theta=0$ and $\theta=1$ coincide, which is a direct consequence of gauge invariance. Thus the spectrum has the cylinder topology where a given energy level is connected to another level according to some specific rule  \cite{remark7}.

One can easily deduce this rule by analyzing the level pattern in the low-energy region, Fig.~\ref{fig12} (left). Indeed, considering $\theta=0$ and appealing to the effective mass approximation we can assign the following quantum numbers to the depicted energy levels: $(n_r,n_L)=(0,0), (0,1), (0,2),\ldots, (0,21)$. The level-spacings in this regime are described by eq.(\ref{b3}). The 22nd energy level has quantum numbers $(1,-1)$ (it is the lowest-energy state belonging to the first Landau level) and next levels $(0,22), (1,0), (0,23), (1,1), (0,24), (1,2),\ldots$. 
When $\theta$ is increased by unity every state is seen to acquire exactly one quantum of angular momentum, as it is expected on the basis of the effective mass approximation. Moreover this is in agreement with our expectation that low-energy states are topologically non-trivial.

This rule can also  be applied  to the medium energy region, however only to the transporting states. For the transporting states, 
and only for those, angular momentum is an approximate quantum number. This is exemplified in the middle panel in Fig.~\ref{fig12},  where the transporting (topologically non-trivial) states coexist with chaotic counter-transporting (topologically trivial) states. One should remark though that the statement `every transporting state acquires exactly one quantum of 
angular momentum'  is valid only if we ignore avoided crossings. The presence of avoided crossings, which reflect hybridization of the transporting and counter-transporting states in the quantum case, has important consequences for the excitation dynamics of the system considered in the next subsection. 

As can be seen from the right panel of Fig.~\ref{fig12} there is just one state that,  upon flux insertion, connects to one
with a larger angular momentum  in the high energy region \footnote{We would like to remind the reader at this point that the high-energy region is entered through a crossover rather than a sharp transition. In the right-most panel of FIG. \ref{fig12}, although an energy regime below the critical energy $E_{\text{cr}} \approx 11 J$ is shown, already most of the states show signatures characteristic for high energies.}. This is a signature that most of the states in this energy region are non-transporting and topologically trivial.

In conclusion, we find that in the low-energy sector of our system, all states are topologically non-trivial because they generate a non-trivial Hall current when a  flux quantum is introduced to the system in the spirit of Laughlin. All states in this regime are transporting. In the high-energy sector in contrast, the system is topologically trivial with no Hall current being supported and no transporting states. In the medium energy regime we find a mixture of topologically trivial (and non transporting) and non-trivial (transporting) states.

\subsection{Excitation dynamics}

\begin{figure}[t]
\includegraphics[width=0.6\columnwidth]{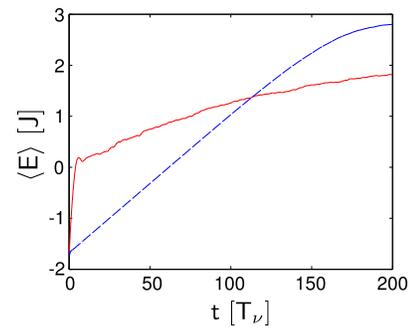}
\caption{(Color online) Mean energy (\ref{d4}) as the function of time for $J=1$, $\nu=0.1$, (blue) dashed line, and $\nu=-0.1$,  (red) solid line. The system parameters are $\alpha=0.1$ and $\gamma=0.018$.}
\label{fig13}
\end{figure}

Let us now consider excitations of the system from its ground state by a time-dependent flux $\theta=\nu t$. 
If $\nu$ is  larger than the characteristic gap of avoided crossings but smaller than the encircling frequency $\Omega$ the 
system is expected to diabatically  (i.e., ignoring tiny avoided crossings) follow the instantaneous energy level and, hence, the mean energy
\begin{equation}
\label{d4}
E(t)=\langle \psi(t)|\widehat{H}(\theta=\nu t)|\psi(t)\rangle
\end{equation}
should monotonically grow in time. For the parameters of Fig.~\ref{fig12} and $\nu=0.1 J$ the average energy, eq. (\ref{d4}), 
is shown in Fig.~\ref{fig13} by the blue dashed line. It is seen that for finite times $E(t)$ indeed shows the expected linear increase which
for longer times however saturates. In the following we discuss these two regimes separately. 

\begin{figure}[t]
\includegraphics[width=0.8\columnwidth]{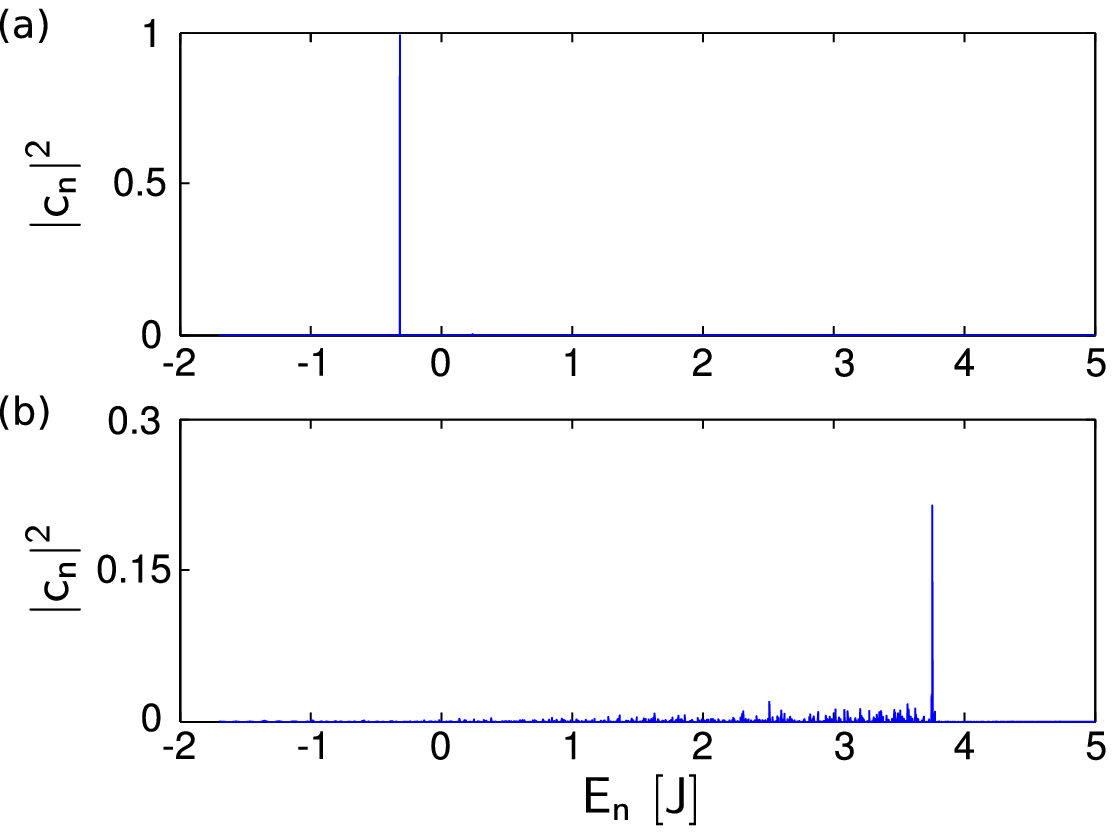}
\caption{Expansion coefficients $c_n$ for the wave function $\psi(t)$ in the instantaneous basis at $t=50 T_\nu$ and $t=200 T_\nu$ ($\nu=0.1 J$).}
\label{fig15}
\end{figure}

\begin{figure}[htb]
\includegraphics[width=0.9\columnwidth]{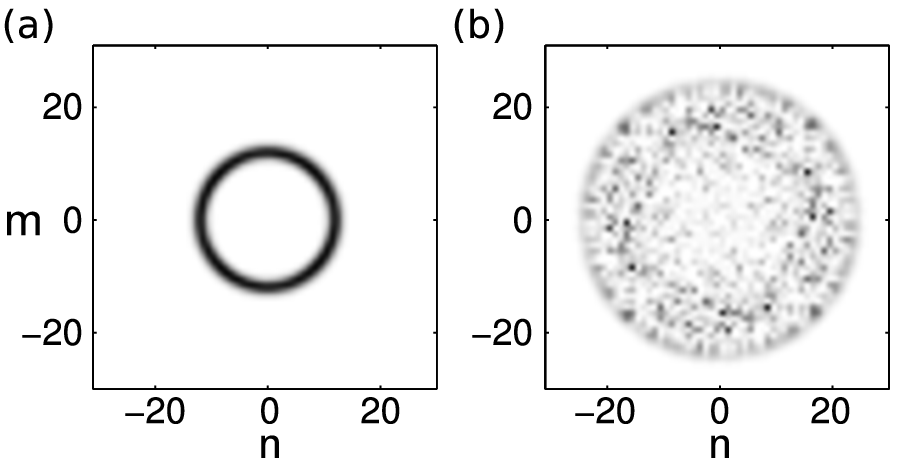}
\caption{Population of the lattice site $|\psi_{n,m}(t)|^2$ at $t=50 T_\nu$ and $t=200 T_\nu$ ($\nu=0.1 J$).}
\label{fig14}
\end{figure}

The linear regime corresponds to diabatic Landau-Zener transitions at the avoided crossing. This is confirmed 
by considering the wave function in the instantaneous basis of the Hamiltonian (\ref{d3}), 
\begin{equation}
\label{d5}
|\psi(t)\rangle =\sum_n c_n(t) |\Psi_n(\theta=\nu t)\rangle \;.
\end{equation}
The upper panel in Fig.~\ref{fig15} shows the expansion coefficients $c_n$ at $t=50T_\nu$, ($T_\nu = 2\pi/\nu$). It can be seen 
that only one level of the Hamiltonian $\widehat{H}(\theta)$ is populated. In real space the ground-state wave function, 
which resembles a wide Gaussian, evolves into a ring with the radius increasing in time, see Fig.~\ref{fig14}(a). Notice that there 
is no `population' inside the ring which is another indication of the diabatic regime.

When the ring is approximately two times larger than in Fig.~\ref{fig14}(a) we observe population of the lattice sites inside the ring. This indicates that the Landau-Zener transitions at the avoided crossings are no longer diabatic  and, hence,  other states of $\widehat{H}(\theta)$ get populated. Finally, for even larger times  the ring fades and population is redistributed over the circle of a finite radius, see Fig.~\ref{fig14}(b). This is the beginning of a self-thermalization process, where  the mean energy $E(t)$ saturates. Remarkably, the saturation energy is found to be substantially smaller than the critical energy (\ref{b4}). This is a manifestation of the hybridization between transporting and counter-transporting states -- the system can reach $\langle E(t)\rangle \simeq  E_{\rm cr}$ only in the classical limit $\alpha,\gamma\rightarrow 0$ ($\alpha^2/\gamma=\text{const}$), where hybridization is absent.

\begin{figure}
\includegraphics[width=0.8\columnwidth]{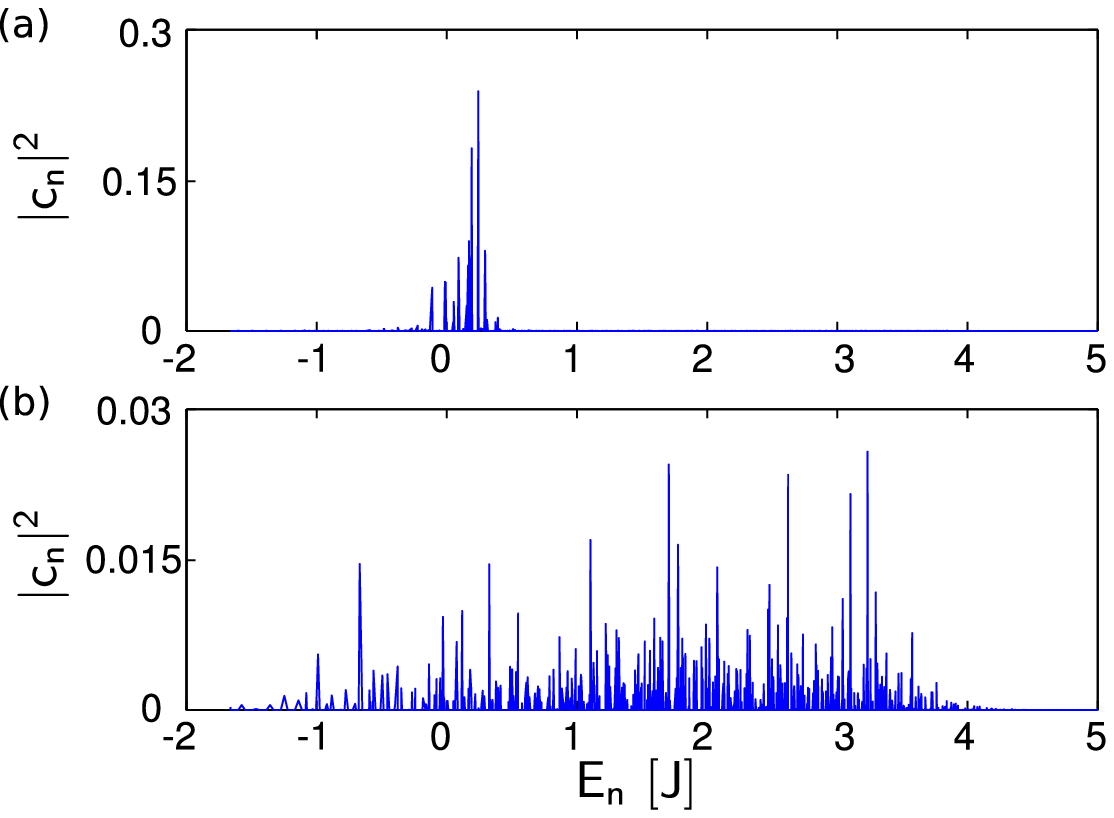}
\caption{(Color online) Expansion coefficients $c_n$ for the wave function $\psi(t)$ in the instantaneous basis at $t=5 T_\nu$ and $t=200 T_\nu$ ($\nu=-0.1 J$).}
\label{fig16}
\end{figure}

\begin{figure}
\includegraphics[width=0.9\columnwidth]{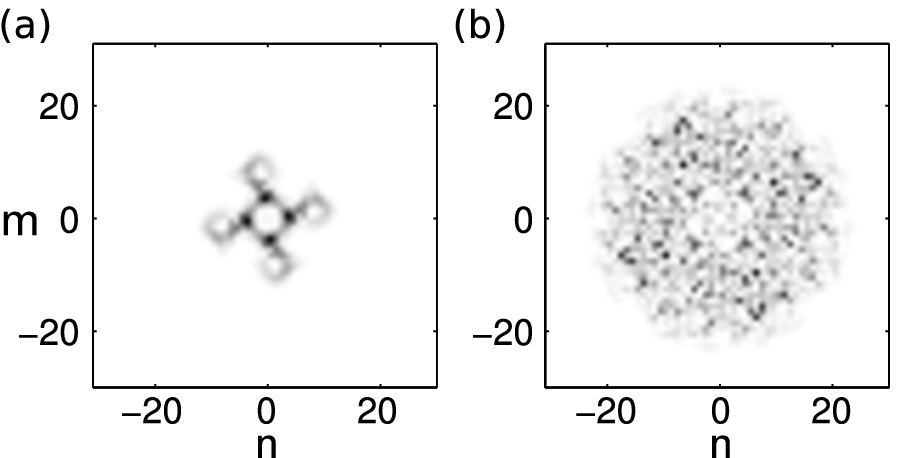}
\caption{Population of the lattice site $|\psi_{n,m}(t)|^2$ at $t=5 T_\nu$ and $t=200 T_\nu$ ($\nu=-0.1 J$).}
\label{fig17}
\end{figure}

We conclude this section by considering the case of negative $\nu$. For negative $\nu$  the time-dependent flux induces a counter-clockwise `circular electric field'  which  is expected to excite the counter-transporting states. The mean energy of the system for negative $\nu=-0.1 J$ is depicted by the red solid line in Fig.~\ref{fig13}.  A rapid initial growth of the energy corresponds to a transient regime where the system follows the energy level connecting the ground state $(n_r,n_L)=(0,0)$ with the state $(n_r,n_L)=(1,-1)$ in the first Landau level, see Fig.~\ref{fig12}(a). After this transient regime the diabatic approximation no longer holds and the subsequent excitation dynamics resembles chaotic diffusion where population is redistributed among many eigenstates of the Hamiltonian (\ref{d3}), including the low-energy states, see Fig.~\ref{fig16}. This hypothesis about chaotic diffusion is further supported by 
the random distribution of population of the lattice cites $|\psi_{n,m}(t)|^2$ at large times, see Fig.~\ref{fig17}(b).

\section{Conclusions}

We analyzed the eigenstates of a quantum particle in a parabolic lattice in the presence
 of a gauge field characterized by the Peierls phase $\alpha$. In a simplified manner the 
 results of this analysis can be summarized as follows.  

In the absence of harmonic confinement and lattice potential the 
eigenstates of the problem are the well-known Landau-states, which are labeled by the 
radial, $n_r$ (Landau level), and the angular momentum, $n_L$, quantum numbers. These 
Landau levels constitute sets of topologically non-trivial states. In the presence 
of a lattice this picture still holds in the low energy regime of small $n_r$ (in particular, $n_r=0$) 
and positive $n_L$ (for positive $\alpha$), where the effective mass approximation is valid. 
We specifically showed that all characteristic features carry over: low-energy states are extended, 
robust against disorder and they produce a quantized Hall current when flux is locally inserted in 
the center of the lattice. Classically, they correspond to regular transporting trajectories.

In the presence of the harmonic confinement (with strength $\gamma$), these Landau-states 
arrange in an equidistant spectrum with frequency separation $\Omega=\gamma/2\pi\alpha$ 
between two states whose angular momentum $n_L$ differs by one unit. When also the
lattice is present, we find three sorts of states: the low-energy sector can still be understood in 
terms of Landau states described above. When the energy increases the discreteness of the 
lattice starts to break the effective mass approximation. In the classical picture, more and more 
transporting trajectories turn into chaotic ones, and the corresponding quantum states can be labeled 
only by their energies. These states are sensitive to disorder and topologically 
trivial, as we deduced from their response to local flux insertion. In the medium energy sector we
furthermore identified a second set of regular transporting trajectories (classical picture), which in the quantum
picture are obtained from the negative-effective mass approximation around the band-maximum. 
In the high-energy sector, above the critical energy $E_{\rm cr}$, all states become localized due to�
Landau-Stark localization. Hence, in this regime there is no transport in the system 
(i.e. the system becomes a trivial insulator) and the eigenstates are topologically trivial.

AK acknowledge the hospitality and financial support of TU Kaiserslatern, where this work was completed. FG is a recipient of a fellowship through the Excellence Initiative (DFG/GSC 266).



\begin{thebibliography}{10}
\bibitem{remark1}
Without a gauge field the quantum particle in 1D and 2D parabolic lattices was discussed earlier in number of paper, see, for example \cite{72,Snoe07} and references therein.
\bibitem{remark2}
In the directions parallel to ${\bf F}$ the localization length of the Landau-Stark scales as $\xi_\parallel\sim J/F$, i.e., the same wave as the localization length of the Wannier-Stark states ($\alpha=0$).
\bibitem{remark7}
In the following discussion we take a convention that we ignore tiny avoided crossing. Without this convention every energy level of the system (\ref{d3}) is a periodic function of $\theta$.
\end{thebibliography}
\end{document}